\documentclass[aps,prl,reprint,groupedaddress]{revtex4-2}

\usepackage{graphicx}
\usepackage{subfig}
\usepackage{chemformula}
\usepackage{amsmath}
\usepackage{tikz, pgfplots}
\usepackage{adjustbox}
\usepackage{amssymb}
\usepackage{siunitx}
\usetikzlibrary{positioning}
\usepackage{caption}
\usepackage{gensymb}
\usepackage[framemethod=tikz]{mdframed}
\usepackage[normalem]{ulem}
\newcommand{\overbar}[1]{\mkern 1.5mu\overline{\mkern-1.5mu#1\mkern-1.5mu}\mkern 1.5mu}
\graphicspath{ {./images/} }
\begin{document}

\title{Comparative analysis of structural, elastic, electronic, phonon, thermal and optical properties of two \ch{Na6Ge2Se6} phases from first principles calculations}

\author{Qi Zhang}
\affiliation{Department of Physics, Missouri University of Science and Technology, Missouri, Rolla 65401, USA}
\author{Amitava Choudhury}
\affiliation{Department of Chemistry, Missouri University of Science and Technology, Missouri, Rolla 65401, USA}
\author{Aleksandr Chernatynskiy}
\affiliation{Department of Physics, Missouri University of Science and Technology, Missouri, Rolla 65401, USA}

\begin{abstract}

The demand for new alkali metal chalcogenide materials is continuously increasing due to their potential applications across various technological fields. Recently, a new compound, \ch{Na6Ge2Se6}, was computationally predicted, representing a new phase distinct from the experimentally observed \ch{Na6Ge2Se6} reported in 1985. Notably, this newly predicted phase displays a lower total energy compared to the previously known experimental phase, as determined by first-principles calculations. In this study, we undertake a thorough comparative analysis of the structural, elastic, electronic, phonon, thermal, and optical properties of these two \ch{Na6Ge2Se6} phases. Our results show that both phases meet mechanical and dynamical stability criteria. The electronic band structure analysis confirms the semiconducting nature of both materials, with a 2.97 eV indirect band gap for the predicted phase and a 2.93 eV direct band gap for the observed phase. Optically, both phases exhibit strong absorption in the ultraviolet region. Thermal properties analysis reveals that the predicted phase is more thermodynamically stable below 907 K, while the observed phase shows greater thermodynamic stability above this temperature.

% has been computationally predicted rencently which has proven to a a new phase to the expeimentally observed \ch{Na6Ge2Se6} in 1985. Remarkably, it exhibits a lower total energy compared to the only experimentally observed \ch{Na6Ge2Se6} phase according to first-principles calculations. In this study, we conduct a comprehensive comparative analysis of the structural, elastic, electronic, phonon, thermal, and optical properties of these two \ch{Na6Ge2Se6} phases. Our findings reveal that they both meet mechanical and dynamical stability criteria with the predicted phase exhibit brittle and ductile for observed phase. The electronic band structures indicate the semiconducting nature of materials with 2.97eV indirect band gap for the predicted phase and 2.93eV direct band gap for the observed phase. Optically, they both shows better absorption in the ultraviolet region. Furthermore, the predicted phase demonstrates greater thermodynamic stability below 105 $K$, whereas the observed phase is more thermodynamically stable above this temperature.

\end{abstract}
\maketitle

\begin{figure*}[t]%
    \centering
    \includegraphics[width=18cm]{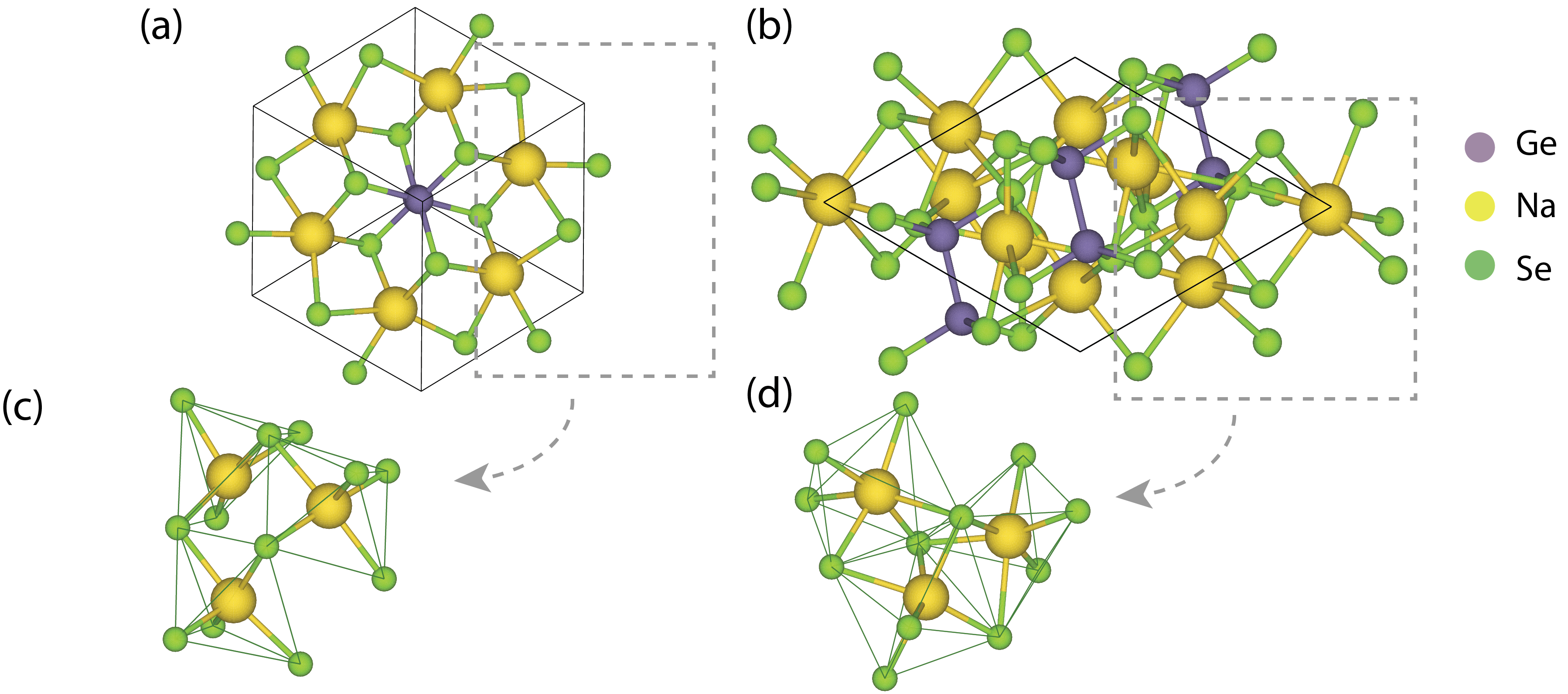}
    \caption{\label{fig:structure}Crystal structures of both \ch{Na6Ge2Se6} phases. (a) Predicted new \ch{Na6Ge2Se6} phase. (b) Observed \ch{Na6Ge2Se6} phase. (c) One Na atom and five Se atoms form a [\ch{NaSe5}] pyramid. (d) One Na atom and six Se atoms form a [\ch{NaSe6}] distorted octahedron.}
\end{figure*}

\section{introduction}

Ternary alkali metal-based chalcogenides are characterized by the general formula ABZ where A signifies an alkali metal (Li, Na, K), B includes main group elements (e.g., Ge, Bi) or transition metals (e.g., Fe, Cu), while Z represents chalcogen elements (e.g., S, Se). These compounds have gained considerable interest recently for their potential applications in energy conversion/storage, optoelectronics, and thermoelectrics. These areas are vital for supporting the advancement of current technologies and industries, especially in addressing the global energy challenge as society aims to transition from fossil fuels to renewable, efficient, and eco-friendly energy solutions. In response to these demands, numerous ABZ compounds have been synthesized and explored.

In the quest for improved energy storage solutions, the focus on ion-battery anodes is crucial due to their significant impact on battery performance. ABZ materials like \ch{LiLnSe2} and \ch{NaFeS2} have shown high storage capacities and efficient energy cycling in recent studies \cite{jia2019ternary,zhang2022erdite}, with \ch{NaFeS2} showcasing excellent cycling stability as a Li-ion battery anode, suggesting potential for enhanced battery lifespan and reliability. Optoelectronically, ABZ compounds such as \ch{NaSbS2} offer strong absorption in the visible spectrum, and their affordability, abundance, and non-toxicity make them attractive for solar energy applications\cite{sun2018eco}. Furthermore, in thermoelectric applications, ABZ materials like \ch{CsAg5Te3} exhibit promising mid-temperature performance with high $ZT$ values\cite{lin2016concerted}, thanks to their intrinsically low thermal conductivity\cite{pei2016multiple,ma2020alpha,ma2019cscu5s3}. Ternary chalcogenides have also played a crucial role in enabling the synthesis of more complex multinary chalcogenides in a rational manner\cite{balijapelly2022building}. Many of these compounds have demonstrated significant technological potential. For instance, outstanding electric thermal properties have been observed in materials like $\text{Na}_{0.95}\text{Pb}_{19}\text{SbTe}_{22}$, achieving a $ZT$ value greater than 1 across one of the broadest temperature ranges (475 to 650 K) reported for any single material\cite{poudeu2006high}. Additionally, a study on $\text{Na}_{1-x}\text{K}_x\text{AsQ}_2$ $(\text{Q}=\text{S,Se})$ highlighted its potential applications in signal processing and data transmission, attributed to its nonlinear optical properties\cite{iyer2021structure}.

Although many ABZ compounds have been synthesised, the existing collection of these ternary ABZ compounds still lacks many possible combinations. The process of discovering new materials through purely experimental means is often prolonged, as the growth and characterization of their properties can be time-consuming. With the increase of  computational power, simulations are increasingly employed in material discovery, for both screening of the properties of already known materials as well as prediction of the new compounds. Recently, a new structure of the chalcogenide, \ch{Na6Ge2Se6} was computationally predicted using simulated annealing and first-principles methods\cite{site}. This compound represents a new phase distinct from the experimentally observed compound reported in 1985 with the same composition (\ch{Na6Ge2Se6})\cite{eisenmann1985oligoselenidogermanate}. The latter has been utilized recently in synthesizing a potential nonlinear optical material, \ch{Na8Mn2(Ge2Se6)2}\cite{balijapelly2022building}. Notably, the newly predicted \ch{Na6Ge2Se6} phase shows a lower total energy compared to the observed phase from first-principles calculations, suggesting it as a potential meta-stable state of \ch{Na6Ge2Se6} that has not been reported before. Furthermore, the observed \ch{Na6Ge2Se6} phase has received little attention, with no comprehensive analysis conducted to the best of our knowledge. This gap in research has motivated us to perform a comparative analysis of the structural, elastic, electronic, phonon, thermal, and optical properties of these two compounds, which could serve as a valuable reference for future studies.

\section{Computational methodology}

In this study, we examined the ground state properties of both predicted and observed \ch{Na6Ge2Se6} phases using density functional theory (DFT) as implemented in the Vienna Ab initio Simulation Package (VASP)\cite{kresse1996efficiency,kresse1996efficient}. This involved solving the Kohn-Sham equations\cite{kohn1965self} to determine the ground state energies of a crystalline system. Given the close proximity of the ground state energies between the predicted and observed phases, calculations were performed using two distinct approaches for the exchange-correlation potential: the Generalized Gradient Approximation (GGA) within the Perdew-Burke-Ernzerhof framework (GGA-PBE)\cite{perdew1996generalized} and the hybrid functionals from Heyd-Scuseria-Ernzerhof (HSE06)\cite{krukau2006influence}. In both \ch{Na6Ge2Se6} structures,  Integration over the Brillouin zone was executed via the tetrahedron method with Gaussian smearing, employing a $5\times5\times5$ Monkhorst–Pack k-point mesh\cite{monkhorst1976special}. The energy cutoff for the plane wave basis set was set to 600 eV. The Ge $3d^{10}4s^34p^2$, Se $4s^2p^4$, Na $2p^63s^1$ electrons are treated as valence electrons. convergence thresholds were \num{1e-8} eV for the electronic self-consistent loop and \num{1e-7} eV for structure optimization.

Further, we explored the elastic and dynamical properties of optimized structures using first-principle calculations with the GGA-PBE functional within VASP. This analysis allowed for the calculation of elastic constants $C_{ij}$ and several elastic properties, including the bulk modulus $B$, Young's modulus $E$ and shear modulus $G$, along with optical properties such as the dielectric function $\epsilon(\omega)$, absorption coefficient $\alpha(\omega)$ and conductivity $\sigma(\omega)$ using the VASPKIT\cite{wang2021vaspkit} tool. The Phonopy package\cite{phonopy-phono3py-JPCM,phonopy-phono3py-JPSJ} facilitated the calculation of phonon dispersion relations, phonon density of states, and related thermal properties. Considering the tendency of GGA-PBE to underestimate band gaps in semiconductors and insulators, the HSE06 functional\cite{krukau2006influence} was applied to investigate electronic and optical properties. The electronic band structures and density of states for both phases were visualized using the open-source Python library Pymatgen\cite{ong2013python}.

\begin{figure}[h]%
    % \centering
    \includegraphics[width=8cm]{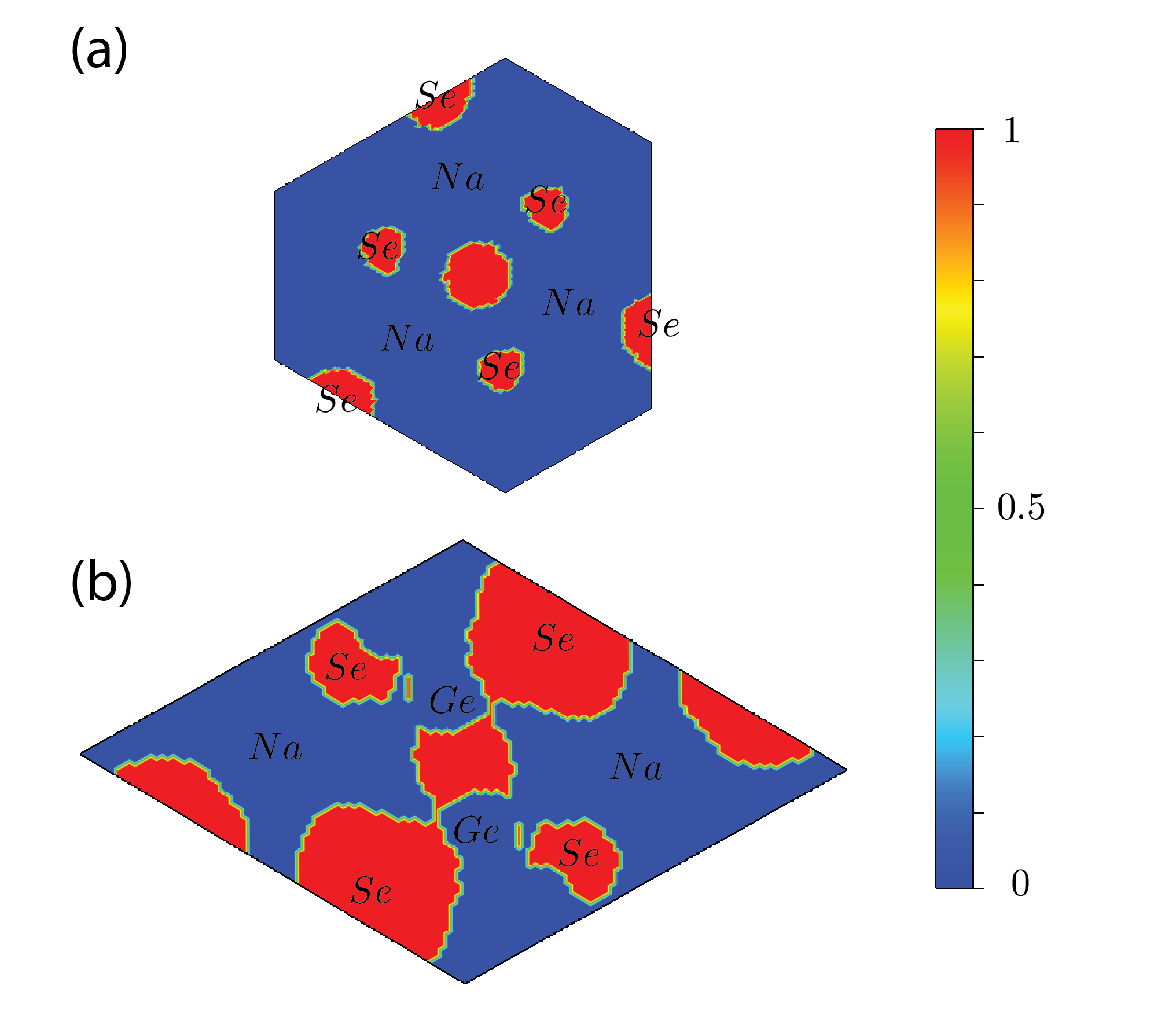}
    \caption{ \label{fig:ELF} Electron localization function (ELF) (a) Predicted \ch{Na6Ge2Se6} phase. (b) Observed \ch{Na6Ge2Se6} phase.}
\end{figure}

\section{Crystal structure}
We analyzed the \ch{Na6Ge2Se6} structures using two different exchange-correlation energy functionals: GGA-PBE and the HSE06 hybrid functional. The calculated energies per atom are presented in Table \ref{tab:crystaldata}. Both PBE and HSE06 functionals indicated lower energies for the predicted \ch{Na6Ge2Se6} phase by 16 meV/atom. To confirm the convergence of our DFT calculations, we conducted two additional structure relaxations with the GGA-PBE functional: one with a 400 $eV$ energy cutoff and a $3\times3\times3$ k-point mesh, and another with a 500 $eV$ energy cutoff and a $4\times4\times4$ k-point mesh. For the predicted phase, both of these settings produced energy per atom within 0.006\% of the most accurate result listed in the Table \ref{tab:crystaldata}. For the observed phase, the difference is even smaller, within 0.002\%. The energy difference between the predicted and observed phases listed above is an order of magnitude larger than 0.06\%, thus numerical accuracy is sufficient to make conclusions about the energy order of two structures.
%the relaxation using a 400 $eV$ energy cutoff and a 3x3x3 k-point mesh resulted in an energy per atom of -3.5838 $eV$, while the relaxation with a 500 $eV$ energy cutoff and a 4x4x4 k-point mesh yielded an energy per atom of -3.58420 $eV$. These results exhibit a negligible difference of less than 0.006\% from the energy values listed in Table \ref{tab:crystaldata}. For the observed phase, the discrepancy is even smaller, with an energy per atom of -3.582055 $eV$ for the 400 $eV$ energy cutoff and 3x3x3 k-point mesh, and -3.582057 $eV$ for the 500 eV energy cutoff and 4x4x4 k-point mesh. This represents a difference of less than 0.002\% from the results in Table \ref{tab:crystaldata}, confirming the convergence of our results.
This leads us to conclude that the predicted \ch{Na6Ge2Se6} phase is energetically more favorable than the observed one at 0 K.

The predicted \ch{Na6Ge2Se6} crystallizes in the space group $R\overbar{3}$ (No.148), whereas the observed phase is found in the space group $P2_1/c$ (No.14). Both structures exhibit identical ethane-like selenide [\ch{Ge2Se6}] units, featuring a Ge-Ge bond length of 2.49 \AA \ for the predicted phase and 2.47 \AA \ for the observed phase, and an average Ge-Se bond length of 2.39 \AA \, as detailed in Table \ref{tab:atom_bonds}. These bond lengths are slightly greater than the originally observed \ch{Na6Ge2Se6} data\cite{eisenmann1985oligoselenidogermanate}, which recorded a Ge-Ge bond length of 2.43 \AA \, and a Ge-Se bond length of 2.33 \AA. The [\ch{Ge2Se6}] ethane-like dimer configuration bears similarity to the [\ch{P2Se6}] in \ch{Mg2P2Se6}\cite{joergens2004kristallstrukturen}, and [\ch{Si2Se6}] in \ch{Na4MgSi2Se6}\cite{wu2015na4mgm2se6}. For both \ch{Na6Ge2Se6} phases, the formed [\ch{Ge2Se6}] dimers are isolated from each other, similar to the cases in \ch{K6Ge2Se6}\cite{eisenmann1984thio} and \ch{Cs6Ge2Se6}\cite{SchlirfDeiseroth}. The predicted \ch{Na6Ge2Se6} phase exhibits cell parameters of $a=11.349$ \AA, $b=11.349$ \AA, $c=10.900$ \AA, $\alpha=90\degree$, $\beta=90\degree$ and $\gamma=120\degree$, while the observed \ch{Na6Ge2Se6} has $a=8.445$ \AA, $b=12.038$ \AA, $c=8.316$ \AA, $\alpha=90\degree$, $\beta=118.877\degree$ and $\gamma=90\degree$ with a smaller unit cell volume. The predicted \ch{Na6Ge2Se6} features only one crystallographically unique Na atom, one Ge atom, and one Se atom, contrasting with the observed \ch{Na6Ge2Se6}, which has three unique Na atoms, one Ge atom, and three Se atoms. In the predicted \ch{Na6Ge2Se6}, each Na atom coordinates with five Se atoms to form a [\ch{NaSe5}] pyramid. This pyramid shares corners with its neighboring [\ch{NaSe5}] pyramids, whereas in the observed phase, each Na atom bonds with six Se atoms to form a [\ch{NaSe6}] slightly distorted octahedron. These octahedra are connected by face-sharing with their neighboring [\ch{NaSe6}] octahedron, as illustrated in Fig. \ref{fig:structure}(c) and Fig. \ref{fig:structure}(d). The electron localization function (ELF) provides a visual representation of valence electron distribution\cite{becke1990simple}, highlighting features that correspond to chemical bonding characteristics. Fig. \ref{fig:ELF} shows that areas near the center of Se atoms and between Ge/Ge atoms exhibit ELF values close to 1.

\begin{table}[t]
\caption{\label{tab:crystaldata}: Crystal data and DFT evaluated energy of predicted and observed \ch{Na6Ge2Se6}}
\begin{ruledtabular}
\begin{tabular}{ccc}
 & Predicted & Observed \\
 \hline & \\[-1.em]
Empirical formula & \ch{Na6Ge2Se6} & \ch{Na6Ge2Se6} \\
Crystal system & Trigonal & Monoclinic \\
Space group & $R\overbar{3}$ & $P2_1/c$ \\
Unit cell & $a=11.349$\AA & $a=8.445$\AA \\
    & $b=11.349$\AA & $b=12.038$\AA \\
    & $c=10.900$\AA & $c=8.316$\AA \\
    & $\alpha=90\degree$ & $\alpha=90\degree$ \\
    & $\beta=90\degree$ & $\beta=118.877\degree$ \\
    & $\gamma=120\degree$ & $\gamma=90\degree$ \\
Volume (\AA$^3$) & $V=405.27$ & $740.20$ \\
$Z$ & 1 & 2 \\
Density($\rho$)& 3.075 $\text{g/cm}^{-3}$& 3.367 $\text{g/cm}^{-3}$\\
Energy (eV/atom) & & \\
GGA-PBE & $-3.598$ & $-3.582$ \\
HSE06 & $-4.268$ & $-4.252$\\
\end{tabular}
\end{ruledtabular}
\end{table}

\begin{table}[t]
\caption{\label{tab:atom_bonds}: Bond distance within the [\ch{Ge2Se6}] block for predicted and observed \ch{Na6Ge2Se6}}
\begin{ruledtabular}
\begin{tabular}{cccc}
\multicolumn{2}{c}{Predicted} & \multicolumn{2}{c}{Observed} \\
  \hline & \\[-1.em]
Ge-Ge & 2.49 \AA & Ge-Ge & 2.47 \AA \\
Ge-Se & 2.39 \AA& Ge-Se & 2.38 \AA\\
\end{tabular}
\end{ruledtabular}
\end{table}

\section{Elastic properties}

\begin{figure*}[t]%
    \centering
    \includegraphics[width=18cm]{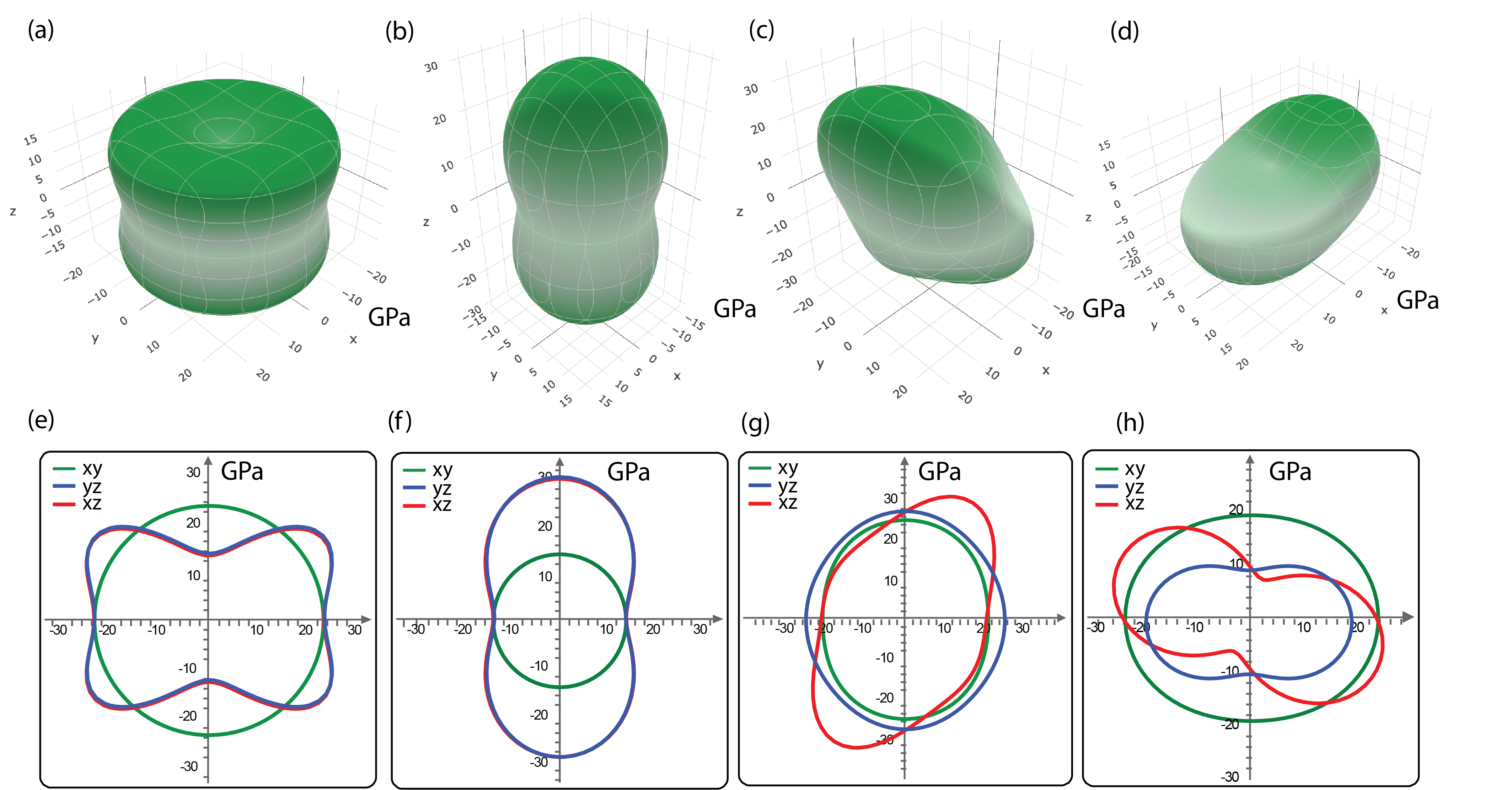}
    \caption{\label{fig:3d}3D representations of the spatial dependence of the Young’s modulus and linear compressibility for (a) and (b) the predicted \ch{Na6Ge2Se6} phase, and (c) and (d) the observed \ch{Na6Ge2Se6} phase. Their cross-sections in the x-, y-, and z-directions are shown below.}
\end{figure*}

Analyzing the elastic properties offers insights into the mechanical behavior of compounds, including their stability, ductility, and brittleness. Such information is especially valuable in industrial and device manufacturing sectors, enhancing our understanding of the forces acting in solids. To calculate elastic constants, we employed the energy-strain method\cite{le2001symmetry}, as facilitated by VASPKIT\cite{wang2021vaspkit} based on DFT calculations with GGA-PBE functional.

For the predicted \ch{Na6Ge2Se6} phase, which crystallizes in the trigonal crystal system, There are five independent
elastic constants: $C_{11}$, $C_{33}$, $C_{44}$, $C_{12}$, $C_{13}$ with the added relation
\begin{equation}
C_{66} = (C_{11} - C_{12})/2
\end{equation}
This contrasts with its observed counterpart, which crystallizes in monoclinic crystal system with lower symmetry that has thirteen independent elastic constants: $C_{11}$, $C_{12}$, $C_{13}$, $C_{15}$, $C_{22}$, $C_{23}$, $C_{25}$, $C_{33}$, $C_{35}$ , $C_{44}$, $C_{46}$ ,$C_{55}$ and $C_{66}$. The configurations of these constants constitute the elastic matrices for each respective phase as follows:
\[
C_{predicted} =
\begin{pmatrix}
    32.06 & 10.86 & 13.31 & \cdot & \cdot & \cdot\\
    \cdot & 32.06 & 13.31 & \cdot & \cdot & \cdot\\
    \cdot & \cdot & 21.06 & \cdot & \cdot & \cdot\\
    \cdot & \cdot & \cdot & 12.09 & \cdot & \cdot\\
    \cdot & \cdot & \cdot & \cdot & 12.09 & \cdot\\
    \cdot & \cdot & \cdot & \cdot & \cdot & 10.60
\end{pmatrix}
\]
\[
C_{observed} =
\begin{pmatrix}
    28.28 & 10.82 & 15.82 & \cdot & 2.45 & \cdot\\
    \cdot & 32.08 & 15.21 & \cdot & 1.55 & \cdot\\
    \cdot & \cdot & 40.55 & \cdot & 4.76 & \cdot\\
    \cdot & \cdot & \cdot & 10.43 & \cdot & 2.81\\
    \cdot & \cdot & \cdot & \cdot & 10.67 & \cdot\\
    \cdot & \cdot & \cdot & \cdot & \cdot & 10.75
\end{pmatrix}
\]

\begin{table}[t]
\caption{\label{tab:mechanical}: Calculated mechanical properties of polycrystal for both predicted and observed \ch{Na6Ge2Se6} (all in $GPa$)}
\begin{ruledtabular}
\begin{tabular}{ccccccc}
& $B$ & $E$ & $G$ & $\nu$ & $B/G$ & $\zeta$ \\
 \hline & \\[-1.em]
Predicted & 17.543 & 24.407 & 9.623 & 0.268 & 1.82 & 0.59\\
 % \hline & \\[-1.em]
Observed & 19.770 & 25.731 & 10.027 & 0.283 & 1.97 & 0.65\\
\end{tabular}
\end{ruledtabular}
\end{table}

\begin{figure*}[t]%
    \centering
    \includegraphics[width=18cm]{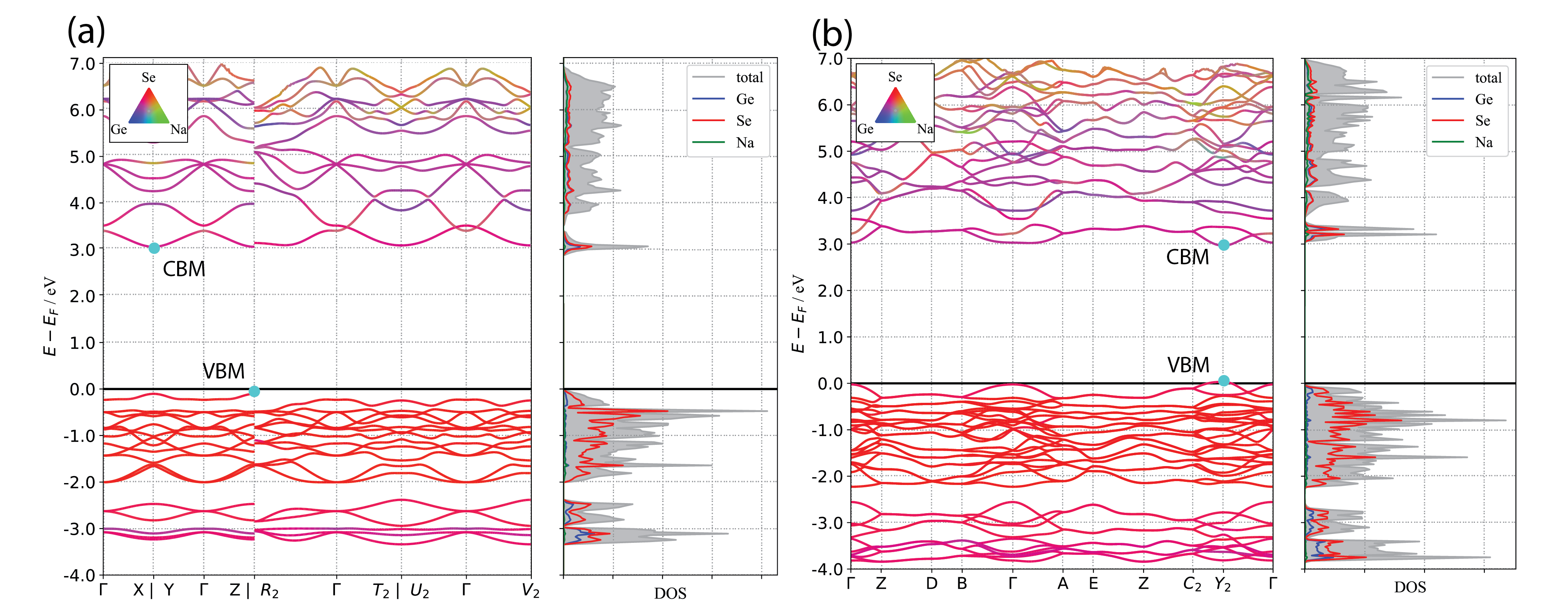}
    \caption{\label{fig:dos}Electronic band structures (BS) and density of states (DOS) of both \ch{Na6Ge2Se6} compounds. (a) BS and DOS for predicted \ch{Na6Ge2Se6} phase. (b) BS and PDOS for observed \ch{Na6Ge2Se6} phase}
\end{figure*}

These elastic constants are pivotal for assessing other elastic properties. A crystal must satisfy the Born-Huang criteria\cite{mouhat2014necessary} to be deemed mechanically stable. For the predicted structure, the necessary and sufficient Born-Huang criteria include:
\begin{equation}
C_{11} > |C_{12}|, C_{44} > 0, C_{66} > 0
\end{equation}
\begin{equation}
2C_{13}^2 < C_{33}(C_{11} + C_{12})
\end{equation}
% \begin{equation}
% C_{11} > 0, C_{44} > 0, C_{55} > 0, C_{66} > 0
% \end{equation}
% % \begin{equation}
% % C_{44} > 0
% % \end{equation}
% % \begin{equation}
% % C_{55} > 0
% % \end{equation}
% % \begin{equation}
% % C_{66} > 0
% % \end{equation}
% \begin{equation}
% C_{11}C_{22} > {C_{12}}^2
% \end{equation}
% \begin{equation}
% \begin{split}
% &C_{11}C_{22}C_{33} + 2C_{12}C_{13}C_{23} \\ & - C_{11}{C_{23}}^2 - C_{22}{C_{13}}^2 - C_{33}{C_{12}}^2 > 0
% \end{split}
% \end{equation}
% \[C_{11} > 0, C_{11}C_{22} > {C_{12}}^2\]
% \[C_{11}C_{22}C_{33} + 2C_{12}C_{13}C_{23} - C_{11}{C_{23}}^2 - C_{22}{C_{13}}^2 - C_{33}{C_{12}}^2 > 0\]
% \[C_{44} > 0, C_{55} > 0, C_{66} > 0\]
Due to the intricacy of these criteria for the monoclinic system, their detailed equations are not listed here. Crucially, the calculated elastic constants for both \ch{Na6Ge2Se6} phases fulfill their respective stability criteria, indicating mechanical stability for both the predicted and observed phases. 

Constants $C_{11}$, $C_{22}$ and $C_{33}$ quantify the response to uniaxial strain along the three principal axes. The differences between these constants measure the extent of anisotropy in linear compressibility. From matrix $C_{predicted}$ and $C_{observed}$, we can see that for predicted phase, $C_{11}=C_{22}>C_{33}$ which means linear compressibility is isotropic in the x- and y-directions and greater than in the z-direction. 
For the observed phase, $C_{33}>C_{22}>C_{11}$, indicating that the observed \ch{Na6Ge2Se6} phase is characterized by anisotropic linear compressibility. While both phases have similar resistance to deformation along the x- and y-directions, the observed phase is significantly stiffer in the z-direction.

On the other hand, $C_{44}$, $C_{55}$ and $C_{66}$ assess the material’s resistance to shear deformation about these axes. The elastic constant matrix for the predicted \ch{Na6Ge2Se6} phase exhibits $C_{44}=C_{55}$ and slightly smaller $C_{66}$, indicating its shear modulus is weakly anisotropic. For this phase, the resistance to shear deformation is isotropic about the x- and y-directions and greater than that in the z-direction. Comparing both phases, the observed phase shows slightly higher resistance to shear deformation about the principal axes.

The off-diagonal terms indicate coupling between different deformation modes. For instance, $C_{12}$ correlates the stress applied in the x-direction to the resultant strain in the y-direction. Analysis of the matrices suggests that both phases have a similar response to deformation in the y-direction when stress is applied in the x-direction. However, the predicted phase shows greater shear strain in the z-direction induced by stress in both the x- and y-directions.

\begin{figure*}[t]%
    \centering
    \includegraphics[width=18cm]{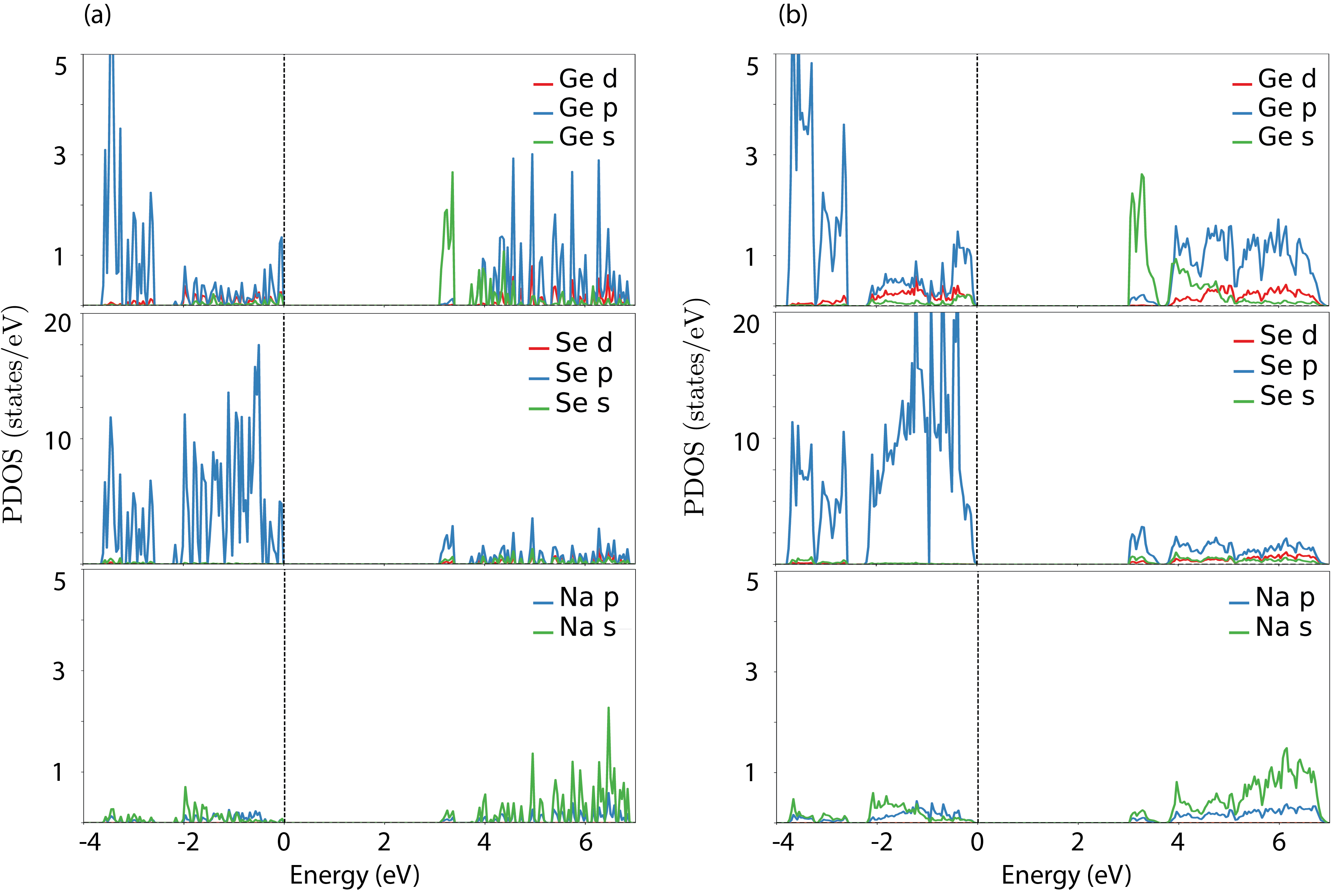}
    \caption{\label{fig:pdos}The partial density of states (PDOS) on orbitals and elements for (a) the predicted \ch{Na6Ge2Se6} and (b) the observed \ch{Na6Ge2Se6}.}
\end{figure*}

\begin{figure*}[t]%
    \centering
    \includegraphics[width=18cm]{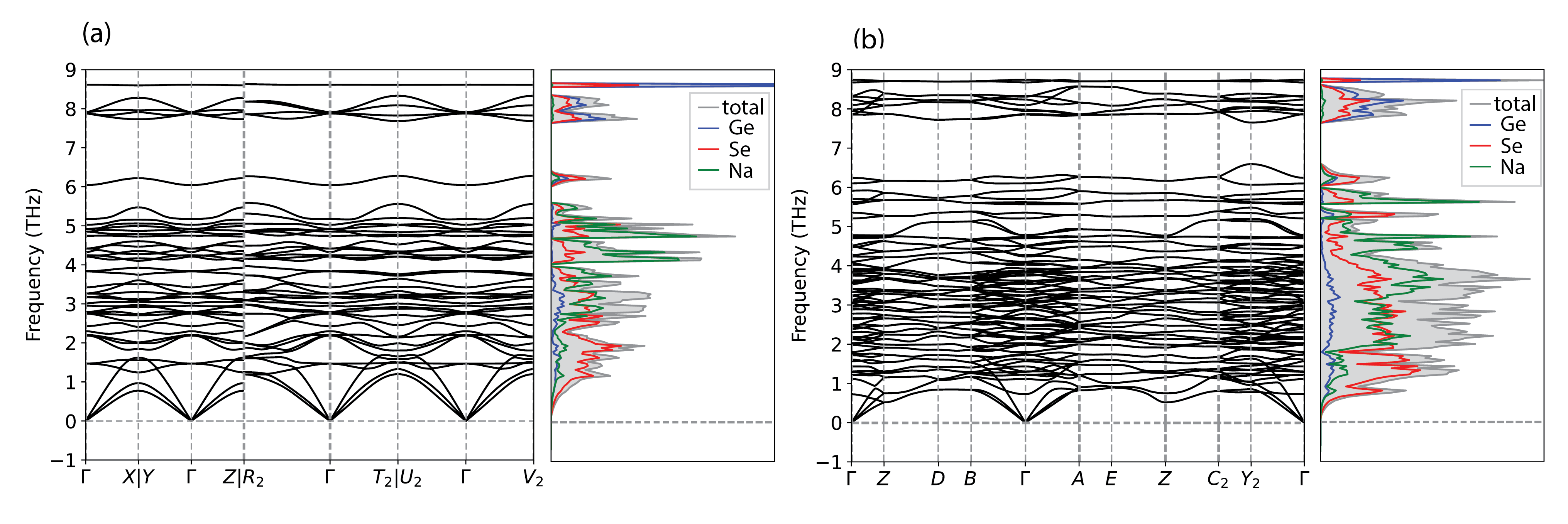}
    \caption{\label{fig:phon_dos}Phonon density of states and band structure of both \ch{Na6Ge2Se6} compounds. (a) Predicted new \ch{Na6Ge2Se6} phase. (b) Observed \ch{Na6Ge2Se6} phase.}
\end{figure*}

Utilizing the independent elastic constants, we can derive key mechanical properties such as the bulk modulus ($B$), shear modulus ($G$), Young's modulus ($E$), and Poisson's ratio ($\nu$) through the following relations, with the calculated values presented in Table \ref{tab:mechanical}:
\begin{equation}
B=\frac{B_V+B_R}{2}
\end{equation}
\begin{equation}
G=\frac{G_V+G_R}{2} 
\end{equation}
\begin{equation}
E=\frac{9BG}{3B+G}
\end{equation}
\begin{equation}
\nu=\frac{3B-2G}{2(3B+G)}
\end{equation}
Here, $B_V$, $B_R$ and $G_V$, $G_R$ represent the Voigt\cite{voigt1910lehrbuch} and Reuss\cite{reuss1929berechnung} values of the bulk modulus and shear modulus respectively. Furthermore, we evaluated Kleinman's parameter ($\zeta$) using VASPKIT\cite{wang2021vaspkit}. 

The bulk and shear moduli offer insights into a crystal's mechanical behavior, indicating its resistance to volumetric and shape deformations, respectively. The data in the Table \ref{tab:mechanical} reveal that the observed \ch{Na6Ge2Se6} phase possesses a greater resistance to deformation. This is expected as the unit-cell volume of the predicted phase is greater than that of the observed phase. Pugh's ratio ($B/G$) serves as a gauge for a material's ductility or brittleness, with a threshold value of 1.75 distinguishing between the two behaviors where the material is claimed to be brittle, if $B/G<1.75$ and classified to be ductile if $B/G>1.75$\cite{pugh1954xcii}. Our findings suggest that both phases exhibit ductility. The stiffness and thermal shock resistance of these materials are inferred from Young's modulus ($E$)\cite{kittel2018introduction}, indicating superior stiffness in the observed phase. Conversely, a lower Young's modulus hints at enhanced thermal shock resistance\cite{karim2018newly} for the predicted phase. Poisson's ratio ($\nu$) sheds light on the nature of bonding forces within a crystal and aids in evaluating its mechanical properties, including stability against shear\cite{ravindran1998density}. Values for the predicted and observed phases are 0.268 and 0.283, respectively, falling within the central force solids' expected range of 0.25 to 0.50\cite{anderson1971elastic}. This metric also predicts the material's ductility or brittleness, classifying those with $\nu>0.26$ as ductile and those below 0.26 as brittle\cite{vaitheeswaran2007elastic,karim2018newly}. Thus, our analysis corroborates both phases' ductility, aligning with conclusions drawn from Pugh's ratio. Lastly, the Kleinman parameter ($\zeta$), varying between 0 and 1, indicates the dominance of bond stretching or bending in resisting external stress\cite{kleinman1962deformation} where the lower value of the Kleinman parameter $\zeta$ indicates a minimal role of bond bending in resisting external stress, whereas a higher value suggests a negligible contribution of bond stretching or contracting to resist externally applied stress. For both \ch{Na6Ge2Se6} phases, a $\zeta>0.5$ indicates a major role for bond bending in its mechanical strength.

Fig. \ref{fig:3d} shows the 3D and 2D representations of Young’s modulus and linear compressibility for both \ch{Na6Ge2Se6} phases. For the predicted phase, the Young’s modulus and linear compressibility in the xy-plane form a circle, as shown in Figures \ref{fig:3d}(e) and (f), suggesting both properties are isotropic within the xy-plane. However, both properties show significant deviation from a circular form within the yz- and xz-planes, indicating that Young’s modulus and linear compressibility are anisotropic overall for the predicted \ch{Na6Ge2Se6}. For the observed phase, Figures \ref{fig:3d}(g) and (h) show that both properties are anisotropic, with the 2D representation of Young’s modulus in the xy- and yz-planes and linear compressibility in the xy-plane more closely resembling a circle. This suggests relative isotropy for both properties within the corresponding planes.

\begin{figure*}[t]%
    \centering
    \includegraphics[width=15cm]{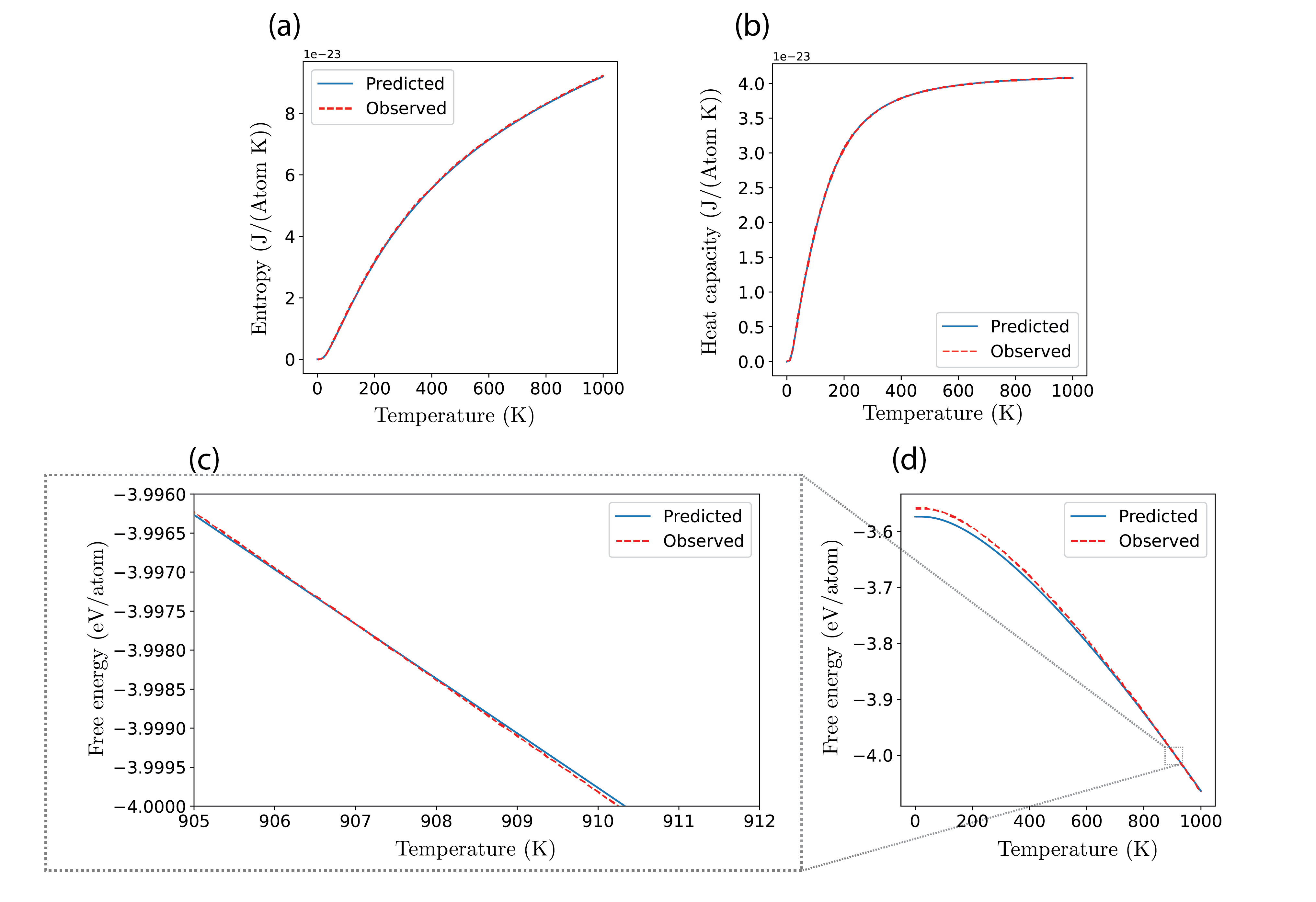}
    \caption{\label{fig:helmholtz}Thermal properties comparison between predicted and observed \ch{Na6Ge2Se6} phases. (a) Entropy. (b) Heat capacity. (c) Helmholtz free energy in the temperature range from 905$K$ to 912$K$. (d) Helmholtz free energy in the temperature range from 0$K$ to 1000$K$}
\end{figure*}

\begin{figure*}[t]%
    \centering
    \includegraphics[width=18cm]{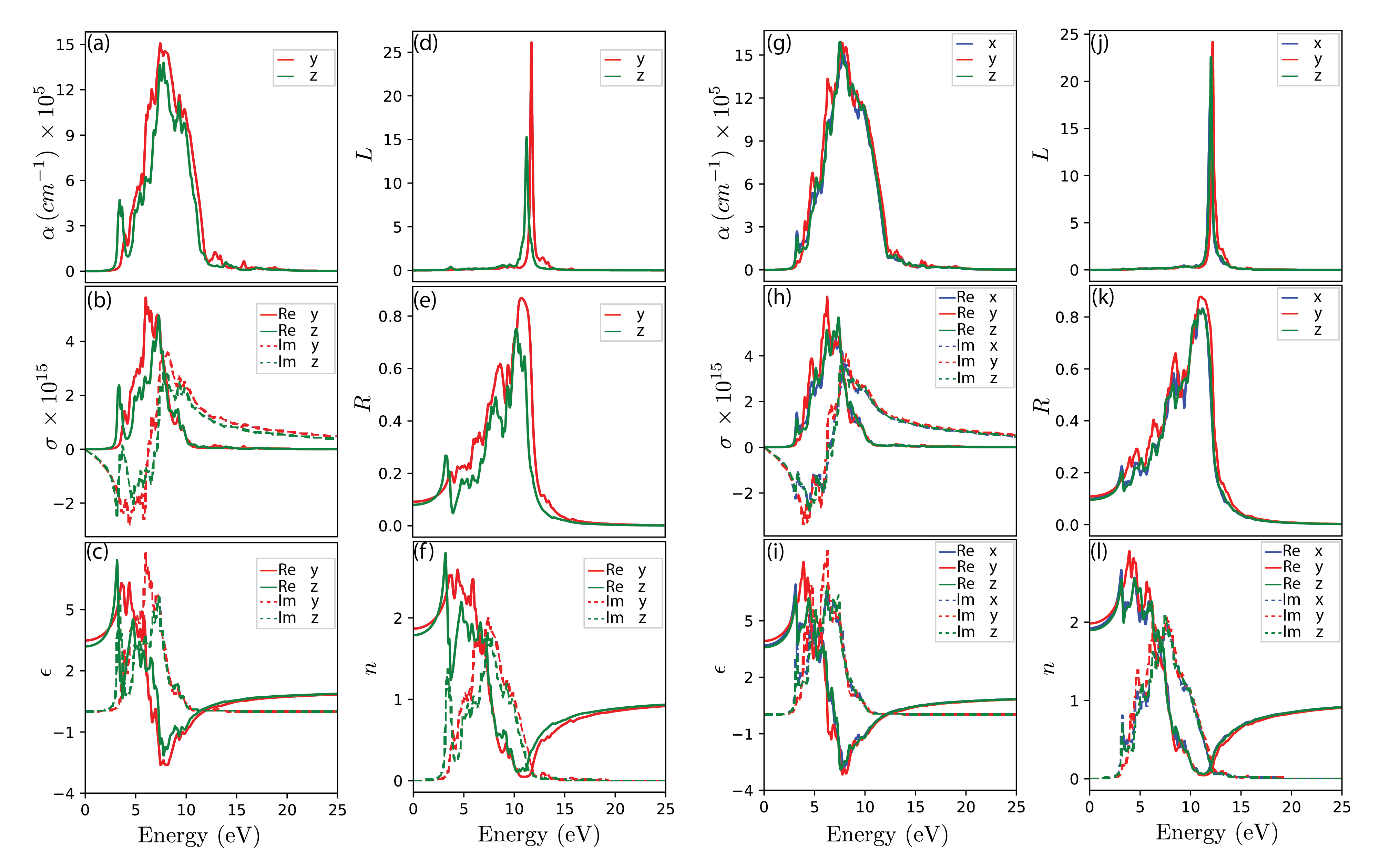}
    \caption{\label{fig:optical}(a)(b)(c)(d)(e)(f) are the optical properties for predicted \ch{Na6Ge2Se6}, (g)(h)(i)(j)(k)(l) are the optical properties for observed \ch{Na6Ge2Se6}. (a)(g): Absorption coefficient $\alpha$, (b)(h): Conductivity $\sigma$, (c)(i): Dielectric function $\epsilon$, (d)(j): Energy-loss function $L$, (e)(k): Reflectivity $R$, (f)(l): Refractive index $n$.
}
\end{figure*}

\section{Electronic Properties}
We present the electronic band structure along with the partial and total density of states (DOS) in Fig. \ref{fig:dos}. These calculations were performed using optimized lattice parameters along high-symmetry directions in the first Brillouin zone. The partial density of states (PDOS) on orbitals and elements are depicted in Fig. \ref{fig:pdos}. The band structures and DOS for both \ch{Na6Ge2Se6} phases were determined using the HSE06 hybrid exchange-correlation functional\cite{krukau2006influence} to ensure enhanced accuracy\cite{heyd2003hybrid}. The energy reference point is set at the top of the valence band. The band structure analysis reveals that the predicted \ch{Na6Ge2Se6} phase possesses an indirect band gap, with the conduction band minimum (CBM) located at the $X$-point and the valence band maximum (VBM) at the $Z$-point, resulting in a band gap of 2.972 eV. In contrast, the observed \ch{Na6Ge2Se6} phase exhibits a direct band gap of 2.932 eV, with both the CBM and VBM situated at the $Y_2$-point. 

From Fig. \ref{fig:pdos}, we observe that both phases share similar orbital and element contributions to the total DOS. Regions close to the Fermi level are predominantly influenced by contributions from Se-p and Ge-s orbitals, while Na’s involvement near the Fermi level is minimal. This observation suggests that the band gaps in both \ch{Na6Ge2Se6} phases are largely attributable to the [\ch{Ge2Se6}] dimers. Within the valence band, ranging from -4 to 0 eV, the primary contributions come from Se-p and Ge-p orbitals in both phases. This trend extends to the lower regions of the conduction band, from 4 to 8 eV, where Na-p displays a slight contribution to the total DOS in this region.

% We present the band structure with optimized lattice parameters along several high symmetry directions in the first brillouin zone are presented, as well as partial and total density of states in Fig. \ref{fig:dos}. We calculated the band structures and dos using HSE06 hybird exchange correlation funtional for both phases for more accurate results. The zero energy is set at the top of the valence band. From the band structure, the predicted \ch{Na6Ge2Se6} phase is identified as a indirect band gap with its conduction band minimum (CBM) at $U_2$-point and the valence band maximum (VBM) at $\gamma$-point with a band gap of 2.972 $eV$. On the other hand, the observed \ch{Na6Ge2Se6} phase has a direct band with the value of the band gap of 2.932 $eV$ which is characterized by the positioning of the CBM and the VBM at the Y2-point. Conversely. The computed total and partial density of states (PDOS) also shown in Fig. \ref{fig:dos}. For both phases, areas proximate to the Fermi level predominantly consist of contribution from Se, along with some from Ge, whereas the Na plays a lesser role near the Fermi level. This suggests that the band gaps in both \ch{Na6Ge2Se6} phases are primarily contributed by the [\ch{Ge2Se6}] dimers. In the valence band, spanning from -4 to 0 eV, the major contributors are Ge and Se for both phases, a pattern that is mirrored at the conduction band's bottom. Additionally, Na contributes modestly to the conduction band's bottom within the 0 to 7 eV range, as depicted in Fig. \ref{fig:dos}.

\section{Phonon and thermal properties}

Investigating phonon properties is fundamental for understanding crystalline materials, as it provides insights into structural stability, phase transitions, and how vibrations influence their thermal behavior. In our study of the phonon dispersion and phonon band structures for both \ch{Na6Ge2Se6} phases, we employed the Phonopy package\cite{phonopy-phono3py-JPCM,phonopy-phono3py-JPSJ} to create a series of $2\times2\times2$ supercell structures with various displacements. We then conducted force calculations using VASP, employing the GGA-PBE exchange-correlation functional. The phonon frequencies and eigenvectors were determined from the dynamical matrices, which were calculated based on the force constants derived using Phonopy.

The phonon dispersion relations along high-symmetry paths for both phases are depicted in Fig. \ref{fig:phon_dos}. The unit cell of the predicted phase comprises 14 atoms, resulting in 42 phonon branches, while the observed phase has 28 atoms, resulting in 82 branches, which include three acoustic and the rest optical branches. The frequency spectrum of these modes spans from 0 to 10 THz, without any discernible gap between the acoustic and optical modes for either phase.
In the lower frequency domain, below 7 THz, Na and Se atoms predominantly contribute to the optical branches. In the higher frequency range, the contributions mainly stem from Ge and Se. This distribution underscores the strong bonding present within the [\ch{Ge2Se6}] dimers, reflecting their structural integrity across both phases of \ch{Na6Ge2Se6}. The lack of negative frequency branches within the dispersion plots affirms the dynamic stability of these phases at zero pressure.

% The phonon dispersion relations along high-symmetry paths for both phases are illustrated in Fig. \ref{fig:pdos_dos}. Given that each unit cell contains 28 atoms for both phases of \ch{Na6Ge2Se6}, there are 84 phonon branches, including three acoustic and 81 optical branches. The frequencies of all acoustic and optical modes range from 0 to 10 $THz$, with no gap observed between the acoustic and optical modes in either phase. Optical branches below 7 $THz$ predominantly arise from Na and Se contributions, while the higher frequency optical branches are mainly due to Ge and Se, highlighting the strong bonding within the [\ch{Ge2Se6}] dimers and indicating their stability in both \ch{Na6Ge2Se6} phases. The absence of negative frequency branches in the dispersion curves confirms the dynamic stability of both phases at zero pressure.

\begin{table}[t]
\caption{\label{tab:debye}: Calculated $v_t$, $v_l$, $v_m$ and $\theta_D$ for both predicted and observed \ch{Na6Ge2Se6}}
\begin{ruledtabular}
\begin{tabular}{ccccc}
& $v_t ($m/s$)$ & $v_l ($m/s$)$ & $v_m ($m/s$)$ & $\theta_D$ ($K$)\\
 \hline & \\[-1.em]
Predicted & 1761.41 & 3129.30 & 1959.71 & 190.0\\
 % \hline & \\[-1.em]
Observed & 1725.51 & 3136.88 & 1923.28 & 191.7\\
  % \hline & \\[-1.em]
\end{tabular}
\end{ruledtabular}
\end{table}

The Debye temperature ($\theta_D$) represents the temperature corresponding to the highest energy vibrational mode in a solid, derived from the equation\cite{anderson1963simplified}:
\begin{equation}
\theta_D=\frac{h}{k_B}\left[\frac{3n}{4{\pi}V_0}\right]^\frac{1}{3}v_m
\end{equation}
Here, $h$ is the Planck’s constant, $k_B$ is the Boltzmann’s constant, $n$ is the number of atoms in the unit cell, $V_0$ is the unit cell's equilibrium volume, and $v_m$ is the material's average speed of sound. The average speed of sound $v_m$ can be determined from the material's mass density ($\rho$), along with its bulk ($B$) and shear modulus ($G$), through the equations for longitudinal ($v_l$) and transverse ($v_t$) sound speeds:
\begin{equation}
v_m=\left[\frac{1}{3}\left(\frac{3}{v_t^3} +\frac{1}{v_l^3}\right)\right]^\frac{1}{3}
\end{equation}
\begin{equation}
v_t=\sqrt{\frac{G}{\rho}}
\end{equation}
\begin{equation}
v_l=\sqrt{\frac{3B+4G}{3\rho}}
\end{equation}
The calculated values for $v_t$, $v_l$, $v_m$ and $\theta_D$ are provided in Table \ref{tab:debye}. These values indicate that the predicted \ch{Na6Ge2Se6} phase exhibits a slower longitudinal speed of sound but a higher transverse one than observed phase, although the values differ only slightly. The Debye temperature of the predicted phase is approximately 0.89\% lower than that of its counterpart.

Phonon calculations offer insights into additional thermal properties such as entropy ($S$), constant-volume heat capacity ($C_v$), and Helmholtz free energy ($F$), which are depicted in Fig. \ref{fig:helmholtz}. The heat capacity curves, as one can see from Fig. \ref{fig:helmholtz}(b), are almost indistinguishable as one expects from similar speeds of sound and Debye temperatures. They show standard Debye theory behavior with the increasing temperature: quantum $~T^3$ at low temperature, saturating to the constant classical Dulong-Petit limit\cite{fitzgerel1960law} above Debye temperature.

The calculation of the Helmholtz free energy ($F$) within the harmonic approximation\cite{FULTZ2010247} facilitates an evaluation of the thermal contributions to the relative stability of two phases, as described by the following equation:
\begin{equation}
\begin{split}
F=&E_{total} + \frac{1}{2}\sum_{q\nu}\hbar\omega(q\nu) \\
&+k_BT\sum_{q\nu}\ln{\left[1-exp(-\hbar\omega(q\nu)/k_BT\right]}
\end{split}
\end{equation}
Here, $E_{total}$ denotes the total energy of the crystal, available in Table \ref{tab:crystaldata}. The summed terms represent the Helmholtz free energy attributable to phonons\cite{togo2015first}, with the initial sum reflecting the zero-point energy (ZPE) that is independent of temperature. The predicted phase's ZPE is approximately 1 meV/atom higher than that of the observed phase. The subsequent sum accounts for the temperature-dependent term referring to the thermally induced occupation of phonon modes. The Helmholtz free energies, plotted in Fig. \ref{fig:helmholtz}(d) for temperatures ranging from 0 to 1000 K, with a closer look at the range of 905 to 912 K in Fig. \ref{fig:helmholtz}(c), illustrate that the predicted phase possesses lower free energy from 0 to approximately 907 K. Since its rate of decrease with temperature is more gradual than that of the observed phase, the free energy of the observed phase becomes lower when the temperature exceeds 907 K. This suggests that the predicted phase is more thermodynamically stable than the observed phase in the temperature range from 0 to 907 K, indicating that this predicted phase is very likely to be experimentally observable. 

\section{Optical properties}
A material's optical behavior is characterized by several energy/frequency-dependent parameters, such as the dielectric function $\epsilon(\omega)$, absorption coefficient $\alpha(\omega)$, conductivity $\sigma(\omega)$, energy-loss function $L(\omega)$, reflectivity $R(\omega)$ and refractive index $n(\omega)$. The outcomes of these calculated properties are depicted in Fig. \ref{fig:optical}, based on formulas integrated within VASPKIT\cite{wang2021vaspkit}.
\begin{equation}
\epsilon(\omega)=\epsilon_1(\omega)+i\epsilon_2(\omega)    
\end{equation}
\begin{equation}
\alpha(\omega)=\frac{\sqrt{2}\omega}{c}\left[\sqrt{\epsilon_1^2+\epsilon_2^2}-\epsilon_1\right]^\frac{1}{2}    
\end{equation}
\begin{equation}
L(\omega)=\frac{\epsilon_2}{\epsilon_1^2+\epsilon_2^2}
\end{equation}
\begin{equation}
n(\omega)=\left[\frac{\sqrt{\epsilon_1^2+\epsilon_2^2}+\epsilon_1}{2}\right]^\frac{1}{2}
\end{equation}
\begin{equation}
R(\omega)=\frac{(n-1)^2+k^2}{(n+1)^2+k^2}
\end{equation}
% \[\alpha(\omega)=\frac{\sqrt{2}\omega}{c}\left[\sqrt{\epsilon_1^2+\epsilon_2^2}-\epsilon_1\right]^\frac{1}{2}, L(\omega)=\frac{\epsilon_2}{\epsilon_1^2+\epsilon_2^2}\]
% \[n(\omega)=\left[\frac{\sqrt{\epsilon_1^2+\epsilon_2^2}+\epsilon_1}{2}\right]^\frac{1}{2},R(\omega)=\frac{(n-1)^2+k^2}{(n+1)^2+k^2}\]
To examine the anisotropic characteristics, we calculated the optical parameters along the three principal axes and presented the optical spectra for incident photon energies up to 25 eV. For the predicted phase, these parameters are isotropic in the x- and y-directions. Therefore, Fig. \ref{fig:optical} displays the optical parameters and the corresponding derived optical properties plotted only in the y- and z-directions for the predicted phase.

The complex dielectric function is essential for describing a material's optical properties, as it serves as the foundation from which other energy-dependent optical constants are derived. We illustrate the dielectric functions for both predicted and observed \ch{Na6Ge2Se6} phases in Fig. \ref{fig:optical}(c) and (i), respectively. The real part of the dielectric function ($\epsilon_1$) initially increases, reaches a peak, then sharply decreases, eventually dropping below zero. The imaginary part ($\epsilon_2$), indicative of dielectric loss, closely aligns with the optical absorption coefficient. From the figure we can see $\epsilon_1$ and $\epsilon_2$ peak in different directions. For the predicted phase, $\epsilon_1$ peaks in z-direction, and $\epsilon_2$ peaks in y-direction. This contrasts with the observed phase, where both $\epsilon_1$ and $\epsilon_2$ peak in the same y-direction. The material's electronic characteristics can be inferred from the absorption coefficient ($\alpha(\omega)$), depicted in Fig. \ref{fig:optical}(a) and (g). Both phases begin to absorb at photon energies of approximately 3.1 eV and 2.9 eV respectively, indicating their optical band gaps. The absorption coefficients in three different directions follow a similar trend, with peaks at 7.4 eV for the predicted phase and 9.5 eV for the observed phase, both within the ultraviolet spectrum. The material's reflection and absorption properties, related to its loss function, are illustrated in Fig. \ref{fig:optical}(e)(k) and (d)(j). The edge trailing of the material's reflection and absorption spectra matches the peak in the loss function, showing a sharp increase around 12 eV for both phases. This spike indicates enhanced absorption and reflection of electromagnetic waves in the ultraviolet region, with the predicted phase exhibiting much notable anisotropy along y- and z-directions with the maximum appearing in y-direction for both phase. Optical conductivity which represents electrical conductivity over specific photon energy ranges is shown in Fig. \ref{fig:optical}(b) and (h). Conductivity sharply increases from zero at photon energies of 3.1 eV and 2.9 eV for the predicted and observed \ch{Na6Ge2Se6} phases respectively, mirroring the optical band gap observed in the absorption curves in Fig. \ref{fig:optical}(a)(g). In summary, the predicted and observed phases display similar optical properties, with the predicted phase exhibiting a higher degree of optical anisotropy in the y- and z-directions, while being isotropic in the x- and y-directions.

\section{Conclusion}
A comprehensive comparison was conducted between two \ch{Na6Ge2Se6} phases: a computationally predicted phase and an experimentally observed phase. This comparison spans their structural, elastic, electronic, phonon, thermal, and optical properties, employing first-principles methods.

The predicted \ch{Na6Ge2Se6} phase is the energetically favored structure at zero pressure and temperature based on first-principles calculations. Both phases feature the identical [\ch{Ge2Se6}] dimer within their crystal structures, with significant differences observed in the bonding environment between Se and Na atoms (five Na atoms bond with one Se atom in the predicted phase, whereas six Na atoms bond with one Se atom in the observed phase).

Regarding elastic properties, both phases meet the criteria for mechanical stability and share similar characteristics. These include their resistance to volumetric and shape deformations, ductility, nature of bonding forces (central force), and bond bending as a major role in resisting external stress. The differences arise from their anisotropic characteristics. While the predicted phase is isotropic in the x- and y-directions, it exhibits greater anisotropy in the y- and z-directions compared to the observed phase.

Electronic structure analyses reveal a closely matched energy band gap between the two phases, primarily contributed by the [\ch{Ge2Se6}] dimers, with only a 1.3\% difference. However, the electronic band structure indicates that the predicted phase exhibits an indirect band gap, while the observed phase has a direct band gap.

The optical behavior between the phases is similarly aligned, with both exhibiting a comparable optical band gap (3.1 eV for the predicted phase and 2.9 eV for the observed phase) and enhanced absorption and reflection of electromagnetic waves in the ultraviolet region. Although the predicted phase is isotropic in the x- and y-directions, it displays a greater degree of optical anisotropy in the y- and z-directions, aligning with the anisotropic nature identified in the elastic properties.

Phonon dispersion analyses affirm the dynamic stability of both phases. Further calculations of thermal properties indicate that the predicted phase is more thermodynamically stable in the temperature range from 0 to 907 K, suggesting a high likelihood of experimental observation.

\section{Acknowledgements}
The authors thank NSF, USA (grant No. DMR1809128) for the funding of this project. The authors also acknowledge the usage of the HPC cluster "Foundry" at Missouri S\&T funded by NSF award OAC-1919789.

\section{data avaiability statement}
The data that support the findings of this study are available upon request.
% \appendix
% \section{Atomic position for both \ch{Na6Ge2Se6}}
% \begin{table}[h]
% \caption{\label{tab:atom_pre}: Fractional atomic coordinates of predicted and observed \ch{Na6Ge2Se6}}
% \begin{ruledtabular}
% \begin{tabular}{cccc}
% Predicted & $x$ & $y$ & $z$ \\
%  \hline & \\[-1.em]
% Na(1) & 0.2187 & 0.5641 & 0.8075 \\
% Na(2) &  0.2426 & 0.0804 & 0.9624  \\
% Na(3) &  0.2508 & 0.6816 & 0.1344  \\
% Na(4) &  0.2536 & 0.8080 & 0.3685  \\
% Na(5) &  0.2586 & 0.4224 & 0.5387  \\
% Na(6) & 0.2790 & 0.9493 & 0.6896  \\
% Ge(1) &  0.1922 & 0.1395 & 0.1807  \\
% Ge(2) &  0.3095 & 0.3485 & 0.3205  \\
% Se(1) &  0.0247 & 0.1329 & 0.7891  \\
% Se(2) &  0.0253 & 0.4184 & 0.3649  \\
% Se(3) &  0.0308 & 0.2946 & 0.0728  \\
% Se(4) &  0.4719 & 0.1966 & 0.4295  \\
% Se(5) &  0.4744 & 0.3791 & 0.7103  \\
% Se(6) &  0.4758 & 0.0691 & 0.1363  \\
%  \hline & \\[-1.em]
%  Observed & $x$ & $y$ & $z$ \\
%   \hline & \\[-1.em]
% Na(1)  &  0.0290 & 0.2291 & 0.7759  \\
% Na(2)  &  0.2750 & 0.5590 & 0.4842  \\
% Na(3)  &  0.4980 & 0.6607 & 0.9776  \\
% Ge   & 0.1599 & 0.5094 & 0.0357  \\
% Se(1)  &  0.1686 & 0.0321 & 0.2608  \\
% Se(2)  &  0.2494 & 0.6971 & 0.1325  \\
% Se(3)  &  0.3377 & 0.1197 & 0.7721  \\

% \end{tabular}
% \end{ruledtabular}
% \end{table}

\bibliography{lib.bib}

\end{document}